\documentclass[aps,prd,twocolumn,showpacs,preprintnumbers,amsmath,amssymb]{revtex4}

\usepackage{graphicx}
\usepackage{dcolumn}
\usepackage{bm}

\begin{document}
\title{$X(3872)$ and the bound state problem of $D^0\bar{D}^{\ast0}$ ($\bar{D}^0{D}^{\ast0}$) in a chiral quark model}

\author{Yan-Rui Liu }
\email{yrliu@ihep.ac.cn}

\author{Zong-Ye Zhang }
\email{zhangzy@ihep.ac.cn}

\affiliation{1. Institute of High Energy Physics, CAS, P.O. Box
918-4, Beijing 100049, China\\
2. Theoretical Physics Center for Science Facilities, CAS, Beijing
100049, China}

\date{\today}

\begin{abstract}

The bound state problem of $D^0\bar{D}^{\ast0}$
($\bar{D}^0{D}^{\ast0}$) is relevant to the molecular interpretation
of the X(3872). We investigated this problem in a chiral quark model
by solving the resonating group method equation. We found the system
is unbound through S-wave $\pi$ and $\sigma$ interactions. The
inclusion of $\rho$ and $\omega$ meson exchanges is helpful to the
formation of a molecule. Because the binding energy relies on the
coupling constants, we cannot draw a definite conclusion whether a
molecular state exists in $D^0\bar{D}^{\ast0}$
($\bar{D}^0{D}^{\ast0}$) system. When moving on to the bottom
counterpart, we obtained an S-wave $B\bar{B}^\ast$ state.

\end{abstract}


\maketitle

\section{introduction}\label{sec1}

In recent years, a number of new charmonium-like states have been
observed \cite{Hmesons,charmonium,XYZ,newHadron}. One of the most
interesting states is the X(3872). The Belle Collaboration first
discovered this state in the $\pi^+\pi^- J/\psi$ channel of $B$
decay in 2003 \cite{3872-first}. Thereafter, CDF \cite{3872-CDF}, D0
\cite{3872-D0}, and BaBar \cite{3872-BaBar} collaborations have
confirmed its existence. The X(3872) is almost on the threshold of
$D^0D^{\ast0}$. Its width is very narrow ($\Gamma< 2.3$ MeV from the
Particle Data Group \cite{PDG}). The measurements from Belle
\cite{3872-angular-Belle} and CDF \cite{3872-angular-CDF} favor the
quantum numbers $J^{PC}=1^{++}$, but $2^{-+}$ have not been ruled
out yet. In the search for a charged X state, BaBar excluded the
isovector hypothesis \cite{3872-charge}.

Experiments have accumulated much information about the decay of the
X(3872). The analysis from CDF \cite{3872-rho-CDF} supports that the
two pions in the channel $X(3872)\to \pi^+\pi^- J/\psi$ come from
the $\rho$ meson. In addition, Belle observed the $3\pi$ decay
$\pi^+\pi^-\pi^0 J/\psi$ and the radiative decay $\gamma J/\psi$
\cite{3872-gamma}. BaBar also reported the evidence of the latter
mode \cite{3872-gamma-Babar}. The measured ratios include
\cite{3872-gamma}
\begin{eqnarray}\label{isov}
\frac{{\cal B}(X(3872)\to \pi^+\pi^-\pi^0 J/\psi)}{{\cal
B}(X(3872)\to \pi^+\pi^- J/\psi)}=1.0\pm 0.4\pm 0.3 ,
\end{eqnarray}
\begin{eqnarray}\label{gammabr-belle}
\frac{{\cal B}[X(3872)\to \gamma J/\psi]}{{\cal B}[X(3872)\to J/\psi
\pi^+\pi^-]}=0.14\pm 0.05
\end{eqnarray}
and \cite{3872-BaBar,3872-gamma-Babar,3872-Babar-ratio}
\begin{eqnarray}\label{gammabr-babar}
\frac{{\cal B}[X(3872)\to \gamma J/\psi]}{{\cal B}[X(3872)\to J/\psi
\pi^+\pi^-]}\approx0.3\,.
\end{eqnarray}
One notes the ratio between the $3\pi$ mode and the dipion mode in
Eq. (\ref{isov}) indicates the large isospin violation when the
X(3872) decays.

Recently, Belle announced a new near-threshold enhancement with
$M=3875.4\pm0.7^{+1.2}_{-2.0}$ MeV in the channel $B\to X(3875)K\to
D^0\bar{D}^{0}\pi^0K$ \cite{3875-Belle}. This state has been
confirmed by BaBar \cite{3875-BaBar}. It is unclear whether or not
these two X states are the same one. If the X(3875) is identical to
the X(3872), there are two more ratios \cite{3875-Belle}
\begin{eqnarray}\label{XtoDpi}
\frac{{\cal B}[X\to D^0\bar{D}^0\pi^0]}{{\cal B}[X\to \pi^+\pi^-
J/\psi]}=8.8^{+3.1}_{-3.6},
\end{eqnarray}
\begin{eqnarray}\label{BtoXK}
\frac{{\cal B}[B^0\to XK^0]}{{\cal B}[B^+\to XK^+]}\approx 1.6\,.
\end{eqnarray}

The exotic properties of the X(3872) have triggered heated
discussions about its nature. The low mass, the extremely narrow
width and the large isospin violation decay are difficult to
understand in the conventional $c\bar{c}$ assignment \cite{ccbar}.
Up to now, there exist many interpretations: a molecular state
\cite{Mole-Close,Mole-Voloshin,Mole-Wong,Mole-Swanson,Mole-Torqvist},
a cusp \cite{3872-cusp}, an S-wave threshold effect
\cite{3872-s-wave}, a hybrid charmonium \cite{3872-hybrid}, a four
quark state \cite{3872-D,3872-tetra}, a vector glueball mixed with
some charmonium components \cite{3872-glueball} and a dynamically
generated resonance \cite{3872-res}. In addition, there are
discussions that the puzzles for the X(3872) may possibly be
resolved in the scheme of mixing
\cite{EichtenLQ,suzuki,chao-3872,Kalashinikova,Pennington}.

The molecular interpretation is the most popular one in
understanding the structure of the X(3872). In fact, the existence
of a loosely bound molecule (deuson) of two heavy mesons has been
proposed long ago \cite{Okun,RGG,Tornqvist}. In such systems, the
contribution from the kinetic term is lowered because of the
presence of the heavy quarks. Since the attraction from the light
quarks is unaffected by the mass of the heavy quark, the formation
of the heavy deuson is possible. According to the calculation in
Ref. \cite{Tornqvist}, several deusons of two bottom mesons should
exist while the predicted deusons of two charmed mesons are close to
the thresholds.

The proximity of the mass of the X(3872) to the threshold of
$D^0\bar{D}^{\ast0}$ ($\bar{D}^0{D}^{\ast0}$) motivated its
molecular interpretation. Numerous discussions have taken place
within this picture
\cite{Petrov,Colangelo,Braaten-ratio,Braaten-3872,Voloshin,Kolck,Hanhart,Ma,Dong}.
The mass, the quantum numbers $J^{PC}$, the isospin violating decay
and the 3$\pi$ decay appear to be naturally understood.

However, the ratios in Eq. (\ref{gammabr-belle}) and Eq.
(\ref{gammabr-babar}) challenged the molecular interpretation. Both
are inconsistent with the molecular picture's prediction which is
around $7\times 10^{-3}$. If the X(3875) and the X(3872) are the
same state, the values in Eq. (\ref{XtoDpi}) and Eq. (\ref{BtoXK})
are also much larger than the theoretical predictions. $\frac{{\cal
B}[X\to D^0\bar{D}^0\pi^0]}{{\cal B}[X\to \pi^+\pi^- J/\psi]}$ from
the molecular assumption is $0.05$ and $\frac{{\cal B}[B^0\to
XK^0]}{{\cal B}[B^+\to XK^+]}$ is less than 0.1
\cite{Braaten-ratio}.

Therefore, whether the molecular picture is correct or not remains
inconclusive. This question is relevant to whether
$D^0\bar{D}^{\ast0}$ ($\bar{D}^0{D}^{\ast0}$) can form a molecule.
Up till now, the dynamical studies of this system are still scarce.
In the calculation of Swanson \cite{Mole-Swanson} and Wong
\cite{Mole-Wong}, binding is possible when the short-range
quark-gluon force is considered. However, the purely molecular
assumption of the X(3872) was questioned in Ref. \cite{suzuki}.

In order to further understand the nature of the X(3872), it is
worthwhile to study dynamically the molecular assumption for the
X(3872) with various methods. In a previous work \cite{liu-3872}, we
have investigated at hadronic level whether the formation of a bound
state of $D^0$ and $\bar{D}^{\ast 0}$ is possible. We found that one
pion and one sigma exchange interactions could not bind the system
to an S-wave molecule. The same framework was also applied to the
newly observed $Z^+(4430)$ \cite{liu-4430}.

In this paper, we reanalyze whether the X(3872) could be an S-wave
$D^0\bar{D}^{\ast0}$ molecule in a different approach. We will study
this system in a chiral constituent quark model and calculate the
binding energy by solving the resonating group method (RGM) equation
\cite{Kamimura,Oka}.

The chiral quark model \cite{SU3CQM} is a useful tool in connecting
QCD theory and the experimental observables, especially for the
light quark systems. This phenomenological model has been quite
successful in reproducing the energies of the baryon ground states,
the binding energy of the deuteron, the nucleon-nucleon ($NN$)
scattering phases and the hyperon-nucleon ($YN$) cross sections. In
this model, the interacting potentials between the two constituent
quarks include the confinement, the one-gluon exchange (OGE) part
and the pseudoscalar and scalar meson exchange part. It has been
controversial whether OGE or vector-meson exchange dominates the
short-range quark-quark interaction in the low-lying baryon states.
Thus the vector meson exchange part has been included in Ref.
\cite{ExCQM}. The model was named as the extended chiral SU(3) quark
model. It was found that the OGE is nearly replaced by the vector
meson exchanges. By solving the RGM equation, the experimental
observables were well reproduced.

Recently, the chiral quark model has been extended to study bound
state problems for the baryon-meson system \cite{BMsystem} and
baryon-antibaryon system \cite{BBbarsys} by solving the RGM
equation. In this work, we will study similar problem for the
$D^0\bar{D}^{\ast0}$ system within this approach.

This paper is organized as follows. In Section \ref{sec2}, we
present the formalism for the calculation. In Section \ref{sec3}, we
give the methods to determine the parameters and their values. Then
in Section \ref{sec4} we show the numerical results, and the last
section gives a summary and discussion.

\section{Formalism}\label{sec2}

\subsection{The molecular picture}

\begin{figure}[htb]
\centering
\includegraphics[scale=0.7]{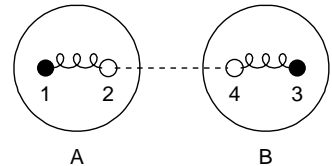}
\caption{The molecule bound by one meson exchange interaction in the
chiral quark model. The solid (empty) dot is the heavy (light) quark
or antiquark. \label{mole} }
\end{figure}

The heavy molecular state bound by the one-meson exchange
interaction in the chiral quark model can be depicted in Fig
\ref{mole}, where $A$ and $B$ are two heavy mesons. The OGE and the
confinement interactions occur inside the color-singlet mesons only.
The interactions between the two clusters are induced by the
one-meson exchange potential between light quarks.

If the X(3872) is really a molecule, the wave function in flavor
space should be \cite{charmonium,liu-3872}
\begin{eqnarray}\label{Xwave-ful}
X(3872)&=&\frac{a_0}{\sqrt{2}}\Big[D^0
\bar{D}^{*0}-{D}^{*0}\bar{D}^0\Big]\nonumber\\
&&+ \frac{a_1}{\sqrt{2}}\Big[D^+ D^{*-}-{D}^{*+}D^-\Big] + \cdots
\end{eqnarray}
where the ellipsis denote other hadronic components. Because of the
large isospin violation in the decay, one expects the first part
dominates with $a_0>>a_1$. In the following calculation, we study
whether $D^0$ and $\bar{D}^{*0}$ may form an S-wave molecule with
the flavor wave function \cite{liu-3872}
\begin{eqnarray}\label{Xwave}
|X_D\rangle=\frac{1}{\sqrt{2}}\Big[|D^0
\bar{D}^{*0}\rangle-|{D}^{*0}\bar{D}^0 \rangle\Big].
\end{eqnarray}

If the answer is yes, this molecular state should lie below the
threshold and identifying the X(3872) as an $X_D$-dominated molecule
is favored. Otherwise, the pure molecular interpretation of the
X(3872) is problematic. We search for an answer by calculating the
binding energy of the system $D^0\bar{D}^{*0}$. Besides the pion and
sigma exchange interactions, the $\rho$ and the $\omega$ exchange
effects are also considered and discussed.

\subsection{Hamiltonian}

The details of the chiral SU(3) quark model can be found in Refs.
\cite{SU3CQM,ExCQM}. Here we just present essential constituents for
the calculation. The Hamiltonian has the form
\begin{eqnarray}\label{ham}
H &=&\sum_{i=1}^4 T_i -T_G + V^{OGE}+V^{conf}+\sum_{M}V^M
\end{eqnarray}
where $T_i$ is the kinetic term of the $i$th quark or antiquark and
$T_G$ is the kinetic energy operator of the center of mass motion.

The potential of the OGE part reads
\begin{eqnarray}
V_{\bar{q}Q}^{OGE}&=&g_qg_Q
\mathbf{F}^c_{\bar{q}}\cdot\mathbf{F}^c_Q\left\{\frac{1}{r}-\frac{\pi}{2}\delta^3(\mathbf{r})\Big[\frac{1}{m_q^2}+\frac{1}{m_Q^2}\right.\nonumber\\
&&\left.+\frac43\frac{1}{m_qm_Q}(\bm{\sigma}_q\cdot\bm{\sigma}_Q)\Big]\right\},
\end{eqnarray}
where $\mathbf{F}^c_{Q}=\frac{\bm{\lambda}}{2}$ for quarks and
$\mathbf{F}^c_{\bar{q}}=-\frac{\bm{\lambda}^\ast}{2}$ for
antiquarks. $m_q$ $(m_Q)$ is the light (heavy) quark mass. The
linear confinement potential is
\begin{eqnarray*}
V_{\bar{q}Q}^{conf}=-4\mathbf{F}^c_{\bar{q}}\cdot\mathbf{F}^c_Q\left(a^c_{aQ}
r +a^{c0}_{qQ}\right).
\end{eqnarray*}
There are similar expressions for $V_{q\bar{Q}}^{OGE}$ and
$V_{q\bar{Q}}^{conf}$.

From Refs. \cite{SU3CQM,ExCQM}, one gets
\begin{eqnarray}
V_{uu}^{\sigma}(\bm{r}_{ij})&=&-C(g_{ch},m_\sigma,\Lambda)X_1(m_\sigma,\Lambda,r_{ij}),\\
V^{\pi_a}(\bm{r}_{ij})&=&C(g_{ch},m_{\pi_a},\Lambda)\frac{m_{\pi_a}^2}{12
m_{q_i}m_{q_j}}
X_2(m_{\pi_a},\Lambda,r_{ij})\nonumber\\
&&\times[\bm{\sigma}(i)\cdot\bm{\sigma}(j)][\tau_a(i)\tau_a(j)],\\
V^{\rho_a}(\bm{r}_{ij})&=&C(g_{chv},m_{\rho_a},\Lambda)\left\{X_1(m_{\rho_a},\Lambda,r_{ij})
+\frac{m_{{\rho_a}}^2}{6
m_{q_i}m_{q_j}}\right.\nonumber\\
&&\times\left(1+\frac{f_{chv}}{g_{chv}}\frac{m_{q_i}+m_{q_j}}{M_N}+
(\frac{f_{chv}}{g_{chv}})^2\frac{m_{q_i}m_{q_j}}{M_N^2}\right)\nonumber\\
&&\times
X_2(m_{{\rho_a}},\Lambda,r_{ij})[\bm{\sigma}(i)\cdot\bm{\sigma}(j)]\Big\}[\tau_a(i)\tau_a(j)],\nonumber\\\\
V_{uu}^\omega(\bm{r}_{ij})
&=&C(g_{chv},m_\omega,\Lambda)\left\{X_1(m_\omega,\Lambda,r_{ij})
+\frac{m_{\omega}^2}{6
m_u^2}\right.\nonumber\\
&&\times\left(1+\frac{f_{chv}}{g_{chv}}\frac{2m_u}{M_N}+
(\frac{f_{chv}}{g_{chv}})^2\frac{m_u^2}{M_N^2}\right)\nonumber\\
&&\times
X_2(m_\omega,\Lambda,r_{ij})[\bm{\sigma}(i)\cdot\bm{\sigma}(j)]\left.\frac{}{}\right\},\\
V_{u\bar{u}}^M&=&G_M V_{uu}^M.
\end{eqnarray}
Where $G_M$ is the G-parity of the exchanged meson and
\begin{eqnarray}
C(g_{ch},m,\Lambda)&=&\frac{g_{ch}^2}{4\pi}\frac{\Lambda^2
m}{\Lambda^2-m^2},\\
X_1(m,\Lambda,r)&=&Y(mr)-\frac{\Lambda}{m}Y(\Lambda r),\\
X_2(m,\Lambda,r)&=&Y(mr)-\left(\frac{\Lambda}{m}\right)^3 Y(\Lambda r),\\
Y(x)&=&\frac{e^{-x}}{x}.
\end{eqnarray}

The tensor term and the spin-orbital term have been omitted in the
potentials since we consider only S-wave interactions. We use the
same cutoff $\Lambda$ for various mesons. Its value is around the
scale of chiral symmetry breaking ($\sim$1 GeV).

\subsection{Bound state problem}

According to the quark cluster model, the wave function of the two
mesons system in coordinate space reads
\begin{eqnarray}
\Psi=\psi_A({\bm \xi}_A)\psi_B({\bm \xi}_B)\chi({\bm R}_{AB})Z({\bm
R}_{cm})
\end{eqnarray}
where $\bm{\xi}_A=\bm{r}_2-\bm{r}_1$ and
$\bm{\xi}_B=\bm{r}_4-\bm{r}_3$ are the internal coordinates of
clusters $A$ and $B$ respectively, ${\bm R}_{AB}$ is the relative
coordinate between the two clusters, and ${\bm R}_{cm}$ is the
center of mass coordinate of the system. $\psi_A(\bm{\xi}_A)$, and
$\psi_B(\bm{\xi}_B)$ are the wave functions of $A$, $B$ and $Z({\bm
R}_{cm})$ represents the center of mass motion wave function of the
system in coordinate space. All of them are treated as Gaussian
functions:
\begin{eqnarray}
\psi_A(\bm{\xi}_A)&=&\left(\frac{m_A\omega}{\pi}\right)^{3/4}e^{-\frac12 m_A\omega\bm{\xi}_A^2},\nonumber\\
\psi_B(\bm{\xi}_B)&=&\left(\frac{m_B\omega}{\pi}\right)^{3/4}e^{-\frac12 m_B\omega\bm{\xi}_B^2}, \nonumber\\
Z({\bm
R}_{cm})&=&\left(\frac{M_{AB}\omega}{\pi}\right)^{3/4}e^{-\frac12
M_{AB}\omega\bm{R}_{cm}^2},
\end{eqnarray}
where $m_A=m_B=\frac{m_q m_Q}{m_q+m_Q}$ is the reduced mass for the
cluster $A$ or $B$ and $M_{AB}=M_A+M_B=2(m_q+m_Q)$ is the total mass
of the two clusters. The universal oscillator frequency $\omega$ is
associated with the width parameter $b_u$ of the up quark through
\begin{equation}
\frac{1}{b_u^2}=m_u \omega \,.
\end{equation}

The unknown relative orbital wave function $\chi(\bm{R}_{AB})$ is
expanded to partial waves
\begin{eqnarray}
\chi(\bm{R}_{AB})&=&\sum_{L=0}^\infty\frac{1}{R_{AB}}\chi^L(R_{AB})Y_{LM}(\hat{\bm
R}_{AB}), \\
\chi^L(R_{AB})&=&\sum_{i=1}^N c_i \,4\pi R_{AB}
\left(\frac{\mu_{AB}\omega}{\pi}\right)^{3/4}
e^{-\frac12 \mu_{AB}\omega(R_{AB}^2+S_i^2)}\nonumber\\
&&\times i_L(\mu_{AB}\omega R_{AB}S_i),
\end{eqnarray}
where $S_i$ ($i$=1, 2,$\cdots$, N) are the generator coordinates,
$\mu_{AB}=\frac12(m_q+m_Q)$ is the reduced mass of the two clusters
and $i_L(x)$ is the modified spherical Bessel function of $L$ order.
The coefficients $c_i$ are to be obtained by solving the
Schr\"odinger equation.

The RGM equation for the bound state problem reads
\begin{eqnarray}\label{RGM-linear}
\sum_{j=1}^N [H_{ij}^L-E N_{ij}^L]c_j=0 \qquad (i=1,\cdots,N)
\end{eqnarray}
where
\begin{eqnarray}
\left\{
\begin{array}{c}H_{ij}^L\\N_{ij}^L\end{array} \right\}&=&\int Y_{LM}^\ast(\hat{\bm S}_i)\left\{
\begin{array}{c}H_{ij}\\
N_{ij}\end{array} \right\}Y_{LM}(\hat{\bm S}_j)d \hat{\bm S}_i d
\hat{\bm S}_j,\nonumber\\
\left\{
\begin{array}{c}H_{ij}\\
N_{ij}\end{array} \right\}&=&\int \Psi({\bm S}_i)\left\{
\begin{array}{c}H\\
1\end{array} \right\}\Psi({\bm S}_j)\prod_{k=1}^4d {\bm r}_k,
\end{eqnarray}
with
\begin{eqnarray}
\Psi({\bm S}_i)&=&\phi_A({\bm \xi}_A)\phi_B({\bm \xi}_B)\chi({\bm
R}_{AB},\bm{S}_i)Z({\bm R}_{cm}),\\
\chi(\bm{R}_{AB},\bm{S}_i)&=&\left(\frac{\mu_{AB}\omega}{\pi}\right)^{3/4}e^{-\frac12
\mu_{AB}\omega(\bm{R}_{AB}-\bm{S}_i)^2}.
\end{eqnarray}
Here $\phi_A$ ($\phi_B$) denotes the total wave function of the
cluster $A$ ($B$), which includes the radial and spin parts.

By solving Eq. (\ref{RGM-linear}), the energy $E$ and the
corresponding relative motion wave function of the system $(c_i)$
are obtained. From the energy $E$, it is easy to derive the binding
energy $E_0=M_{D^0}+M_{D^{\ast0}}-M_{X_D}$. If $E_0$ is negative,
the system would be unbound.

\section{Determining the parameters}\label{sec3}

There are numerous parameters in the Hamiltonian and the wave
functions: $g_q$, $g_Q$, $a_{qQ}^c$, $a_{qQ}^{c0}$, $m_Q$, $m_u$,
$\omega$, $g_{ch}$, $g_{chv}$, $f_{chv}$ and $\Lambda$. The mass of
the phenomenological $\sigma$ meson is also treated as an adjustable
parameter.

One should note, $M_{D^0}$, $M_{D^{\ast0}}$ and $M_{X_D}$ are all
calculated with the Gaussian functions presented in the former
section. The binding energy will be irrelevant to the internal
potentials of the color-singlet mesons because of the cancellation.
That is, the form of the confinement and the values of $g_q$, $g_Q$,
$a_{qQ}^c$ and $a_{qQ}^{c0}$ will not give effects to the numerical
result of $E_0$. This feature can be understood with the effective
potential between the clusters $A$ and $B$ in the generator
coordinate method:
\begin{eqnarray}
V^L(S_i,S_j)=\frac{V^L_{ij}}{N^L_{ij}}-V_{D^0}-V_{D^{\ast0}}.
\end{eqnarray}
One can examine that the parts due to $V^{OGE}$ and $V^{conf}$ of
Eq. (\ref{ham}) are exactly zero. Therefore, we may take any values,
in principle, for these four parameters. In the following
calculation, we deduce $g_Q$, $a_{qQ}^c$ and $a_{qQ}^{c0}$ by
fitting the masses of the ground state mesons $D$, $D^\ast$, $D_s$,
$D_s^\ast$, $J/\psi$ and $\eta_c$ using a least square fit with the
assumption $a_{cu}^c=a_{cs}^c=a_{cc}^c$.

In the determination, we treat $m_u$, $m_s$, $\omega$ (or $b_u$),
$m_Q$, $g_u$ and $g_s$ as inputs. For the up and strange quark
masses, we use the values given in the previous work
\cite{SU3CQM,ExCQM,BMsystem,BBbarsys} $m_u=313$ MeV and $m_s=470$
MeV. The width parameter $b_u=0.5$ fm in the chiral SU(3) quark
model while $b_u=0.45$ fm in the extended chiral SU(3) quark model.
These values have been fitted to reproduce the masses of the ground
state baryons, the binding energy of the deuteron and the $NN$ and
$YN$ scattering observables. To see the effects of this parameter,
we also use a larger value for the width parameter $b_u=0.6$ fm. To
investigate the heavy quark mass dependence, we take several typical
values $m_c=1430$ MeV \cite{zzz}, $m_c=1550$ MeV \cite{Vijande} and
$m_c=1870$ MeV \cite{Semay}. For the coupling constants, we can use
$(g_u,g_s)=(0.886,0.917)$, (0.886,0.755), (0.875,0.920),
(0.237,0.451) or (0.363,0.500) \cite{BMsystem,zhangd}. With these
inputs, one gets sets of fitted values. Selected results are
presented in Table \ref{innerpara}.

\begin{table}[htb]
\centering
\begin{tabular}{ccccccc}\hline
$m_c$ (MeV)&$b_u$ (fm) &$g_u$ &$g_c$ & $a_{uc}^c$ (MeV$^2$) &
$a_{uc}^{c0}$ (MeV)
\\\hline
1430 & 0.45 & 0.237 & 0.718 & 45548 & -166.07\\
     & 0.5 & 0.886 & 0.774 &51320 & -143.36\\
     &0.6 & 0.886 & 1.086 &54343 &-150.96
\\\hline
1550 & 0.45 & 0.886 & 0.642 &43129 &-150.08\\
     & 0.5 & 0.886 & 0.772 &47360 &-152.71 \\
     &0.6 & 0.875 & 1.097 & 51600 &-161.86\\\hline
1870 & 0.45 & 0.363 & 0.771 & 27296 & -187.07\\
     & 0.5 & 0.886 &0.858 & 35445 & -168.59\\
     &0.6 & 0.886 & 1.165 & 42579 & -180.69\\\hline
\end{tabular}
\caption{Fitted parameters for the calculation in the hidden charm
case.}\label{innerpara}
\end{table}

Actually, in the two meson molecule picture (see Fig.1), the meson
exchanges play the dominant role in the energy of the system. The
parameters of this part include the quark-meson coupling constants
and the meson masses. In the chiral quark model, the $\pi$ and
$\sigma$ exchanges have the same coupling constant, named $g_{ch}$,
because of the chiral symmetry requirement. The coupling constant
$g_{ch}$ is fixed through
\begin{eqnarray}
\frac{g_{ch}^2}{4\pi}=\frac{9}{25}\frac{g_{NN\pi}^2}{4\pi}\frac{m_u^2}{m_N^2}
\end{eqnarray}
with $g_{NN\pi}^2/(4\pi)=13.67$ determined experimentally, from
which one has $g_{ch}=2.621$. Thus, when the vector meson
exchanges are not included,the mass of $\sigma$ and the cutoff
mass $\Lambda$ are the only adjustable parameters which can be
fixed in the light quark systems. For the coupling constants of
the vector meson exchange, one can use
$(g_{chv},f_{chv})=$(3.0,0.0) \cite{Riska}, (2.09,5.26),
(2.351,0.0), and (1.972,1.315) \cite{ExCQM}.

To study the effects due to the uncertainty of the mass of the
$\sigma$, we use $m_\sigma$=595 MeV, 535 MeV and 547 MeV
\cite{ExCQM,BMsystem}. For other mesons, we take the masses from the
particle data book \cite{PDG}: $m_{\pi^0}=134.98$ MeV,
$m_{\rho^0}=775.8$ MeV and $m_\omega=782.59$ MeV. In the
calculation, we adopt the cutoff $\Lambda=1000$ MeV, 1100 MeV and
1500 MeV.

In this work, we also calculated the binding energy for the case
of bottom analog where the flavor wave function is
\begin{eqnarray}
|X_B\rangle=\frac{1}{\sqrt{2}}\Big[|B^+ B^{*-}\rangle-|{B}^{*+}B^-
\rangle\Big].
\end{eqnarray}
The procedure to determine the parameters is very similar. Now the
ground state mesons $B$, $B^\ast$, $B_s$, $B_s^\ast$,
${\varUpsilon}(1S)$ and $\eta_b$ are used. As an input, we choose
$m_b$= 4720 MeV which is close to the value in Ref. \cite{zzz2},
$m_b$=5100 MeV \cite{Vijande2} and $m_b$=5259 MeV \cite{Semay}. By
repeating the fitting procedure, one gets sets of parameters. We
present the selected results in Table \ref{innerpara-b}.

\begin{table}[htb]
\centering
\begin{tabular}{ccccccc}\hline
$m_b$ (MeV)&$b_u$ (fm) &$g_u$ &$g_b$ & $a_{ub}^c$ (MeV$^2$) &
$a_{ub}^{c0}$ (MeV)
\\\hline
4720 & 0.45 & 0.886 & 0.897 & 63452 & -149.44\\
     & 0.5 & 0.875 & 1.100 &68609 & -153.70\\
     &0.6 & 0.237 & 1.567 & 66134 &-207.04
\\\hline
5100 & 0.45 & 0.875 & 0.931 &46250 &-177.68\\
     & 0.5 & 0.363 & 1.208 &47881 &-200.74 \\
     &0.6 & 0.875 & 1.551 & 61739 &-194.71\\\hline
5259 & 0.45 & 0.886 & 0.943 & 39073 & -183.09\\
     & 0.5 & 0.363 & 1.218 & 41301 & -210.85\\
     &0.6 & 0.886 & 1.565 & 56580 & -204.13\\\hline
\end{tabular}
\caption{Fitted parameters for the calculation in the hidden bottom
case.}\label{innerpara-b}
\end{table}

\section{Numerical Results}
\label{sec4}

\begin{figure}[htb]
\begin{center}
\includegraphics[angle=270,scale=0.36]{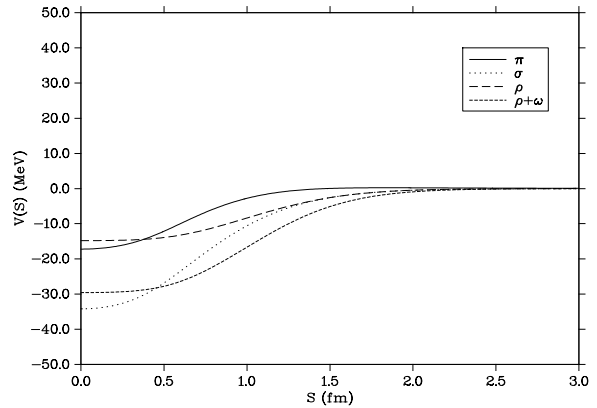}
\caption{The effective potential $V(S)$ for different meson
exchanges. The parameters used are $b_u=0.5$ fm, $m_u=313$ MeV,
$m_c=1870$ MeV, $g_{ch}=2.621$, $g_{chv}=3.0$, $f_{chv}=0.0$,
$m_\sigma=595$ MeV, $m_\pi=134.98$ MeV, $m_\rho=775.8$ MeV,
$m_\omega=782.59$ MeV and $\Lambda=1100$ MeV. Here the line for
$\pi$ corresponds to $V^\pi(S)$, the line for $\sigma$ corresponds
to $V^\sigma(S)$, and so on. }\label{VS}
\end{center}
\end{figure}

Before the numerical evaluation, we first take a look at the
effective potential
\begin{eqnarray}
V(S)=V^{L=0}(S,S)
\end{eqnarray}
where the generator coordinate $S$ can qualitatively describe the
distance between the two clusters. These potentials rely on the
meson exchange part in Eq. (\ref{ham}). We illustrate the
potentials corresponding to various considerations in Fig.
\ref{VS}.

From Fig \ref{VS}, one sees that the interactions from $\pi$,
$\sigma$, $\rho$ and $\omega$ are all attractive \cite{myerr}. The
amplitudes for $\rho$ and $\omega$ exchanges are comparable and
their contributions should not be ignored arbitrarily. We will
consider $\pi$ and $\sigma$ interactions for the moment and then
include the vector meson contributions.

Now we calculate the binding energy for the system
$D^0\bar{D}^{\ast0}$ through solving Eq. (\ref{RGM-linear}). Here we
do not constrain the sets of the parameters with the experimental
data like the studies in Ref. \cite{ExCQM,BMsystem}. After exploring
all possible combinations of the parameters in the former section,
we fail to get a bound state solution. Thus the $D^0\bar{D}^{\ast0}$
system is unbound when we consider only S-wave $\pi$ and $\sigma$
exchange interactions in this framework. This conclusion agrees with
that of Ref. \cite{liu-3872}.

Since the bottom quark is much heavier, the possibility of getting a
bound state in the $B$ meson systems is increased. Our former
dynamical calculation is in favor of the existence of an S-wave
$X_B$ molecule. We also study this case in the present framework.
When applying the evaluation to the bottom analog $B\bar{B}^\ast$,
we get positive binding energies with the parameters in Section
\ref{sec3}. The results are given in Table \ref{E0b}. From that
table, one finds a larger binding energy can be obtained with a
larger $m_b$, a smaller $m_\sigma$, a smaller $b_u$ or a bigger
cutoff $\Lambda$. A deeper bound state should have a smaller
root-mean-square radius $r_{\rm rms}$, which is also illustrated in
Table \ref{E0b}.

\begin{table}[htb]
\centering
\begin{tabular}{ccccc}\hline
$m_b$ (MeV)&$b_u$ (fm) &$m_\sigma$ (MeV) & $E_0$ (MeV) & $r_{\rm
rms}$ (fm)
\\\hline
4720 & 0.45 & 595 & 3.3/3.7/5.0 &1.1/1.1/1.1\\
     &      & 547 & 5.0/5.5/7.0 &1.1/1.1/1.0\\
     &      & 535 & 5.4/6.0/7.5 &1.1/1.1/1.0\\
     & 0.5  & 595 & 2.0/2.3/3.1 &1.3/1.3/1.2\\
     &      & 547 & 3.4/3.8/4.7 &1.2/1.2/1.2\\
     &      & 535 & 3.8/4.2/5.2 &1.2/1.2/1.2\\
     & 0.6  & 595 & 0.5/0.7/1.0 &1.6/1.6/1.5\\
     &      & 547 & 1.5/1.7/2.1 &1.5/1.5/1.5\\
     &      & 535 & 1.8/1.9/2.4 &1.5/1.5/1.4\\\hline
5100 & 0.45 & 595 & 4.2/4.7/6.1 &1.1/1.1/1.0\\
     &      & 547 & 6.0/6.6/8.2 &1.0/1.0/1.0\\
     &      & 535 & 6.5/7.1/8.8 &1.0/1.0/1.0\\
     & 0.5  & 595 & 2.8/3.1/4.0 &1.2/1.2/1.2\\
     &      & 547 & 4.2/4.6/5.7 &1.2/1.2/1.1\\
     &      & 535 & 4.6/5.1/6.2 &1.2/1.1/1.1\\
     & 0.6  & 595 & 1.0/1.2/1.6 &1.5/1.5/1.4\\
     &      & 547 & 2.0/2.2/2.7 &1.4/1.4/1.4\\
     &      & 535 & 2.3/2.5/3.0 &1.4/1.4/1.4\\\hline
5259 & 0.45 & 595 & 4.6/5.1/6.6 &1.1/1.0/1.0\\
     &      & 547 & 6.4/7.0/8.7 &1.0/1.0/1.0\\
     &      & 535 & 6.9/7.6/9.3 &1.0/1.0/1.0\\
     & 0.5  & 595 & 3.1/3.4/4.4 &1.2/1.2/1.1\\
     &      & 547 & 4.6/5.0/6.1 &1.1/1.1/1.1\\
     &      & 535 & 5.0/5.4/6.5 &1.1/1.1/1.1\\
     & 0.6  & 595 & 1.2/1.4/1.8 &1.5/1.5/1.4\\
     &      & 547 & 2.2/2.4/2.9 &1.4/1.4/1.3\\
     &      & 535 & 2.5/2.7/3.3 &1.4/1.4/1.3
\\\hline
\end{tabular}
\caption{Numerical results for the hidden bottom case when $\pi$ and
$\sigma$ exchange potentials are considered. The first
(second,third) value for $E_0$ and $r_{\rm rms}$ corresponds to the
cutoff $\Lambda$=1000 (1100,1500) MeV.}\label{E0b}
\end{table}

To explore additional effects, we move on to include the vector
meson exchanges. We use the parameters to reproduce experimental
data for light quark systems \cite{ExCQM}. The parameters and the
results for $D^0\bar{D}^{\ast0}$ and $B^+\bar{B}^{\ast-}$ are
presented in Table \ref{E0c-vec} and \ref{E0b-vec}, respectively.
For comparison, the solutions without considering vector mesons are
also given. Now a bound state seems to be possible in the
$D^0\bar{D}^{\ast0}$ system. For its bottom analogy, the vector
meson exchange interactions increase the binding energy about 10-20
MeV.

\begin{table}[htb]
\centering
\begin{tabular}{ccccc}\hline
    & $\chi$QM & \multicolumn{2}{c}{Ex. $\chi$QM } \\
    &                 & $f_{chv}/g_{chv}=0$ & $f_{chv}/g_{chv}=2/3$
\\\hline
$b_u$ (fm)& 0.5 & 0.45 & 0.45\\
$m_\sigma$ (MeV)& 595 & 535 & 547\\
$g_{chv}$ &     & 2.351 & 1.972 \\
$E_0$ (MeV)&$\times$&$\times$/$\times$/1.7&$\times$\\
$r_{\rm rms}$ (fm)&$\times$&$\times$/$\times$/1.7&$\times$ \\\hline
$E_0$ (MeV)&$\times$&$\times$/$0.5$/3.0&$\times$\\
$r_{\rm rms}$ (fm)&$\times$&$\times$/$1.8$/1.6&$\times$
\\\hline
\end{tabular}
\caption{Numerical results for the hidden charm case when the vector
meson exchange interactions are also included. Here $m_u=313$ MeV,
$g_{ch}=2.621$ are used. The three values for $E_0$ and $r_{\rm
rms}$ correspond to $m_c=1430$ MeV, 1550 MeV and 1870 MeV in order.
The first (second) $E_0$ and $r_{\rm rms}$ correspond to
$\Lambda$=1100 (1500) MeV. $\times$ indicates the system is
unbound.}\label{E0c-vec}
\end{table}

\begin{table}[htb]
\centering
\begin{tabular}{ccccc}\hline
    & $\chi$QM & \multicolumn{2}{c}{Ex. $\chi$QM } \\
    &                 & $f_{chv}/g_{chv}=0$ & $f_{chv}/g_{chv}=2/3$
\\\hline
$b_u$ (fm)& 0.5 & 0.45 & 0.45\\
$m_\sigma$ (MeV)& 595 & 535 & 547\\
$g_{chv}$ &     & 2.351 & 1.972 \\
$E_0$ (MeV)&2.3/3.1/3.4&17.2/18.7/19.3&11.1/12.3/12.8\\
$r_{\rm rms}$ (fm)&1.3/1.2/1.2&0.9/0.9/0.9&1.1/1.0/0.9 \\
\hline
$E_0$ (MeV)&3.1/4.0/4.4&20.7/22.4/23.1&13.6/15.0/15.6\\
$r_{\rm rms}$ (fm)&1.2/1.2/1.1&0.9/0.8/0.8&1.0/0.9/0.9
\\\hline
\end{tabular}
\caption{Numerical results for the hidden bottom case when the
vector meson exchange interactions are also included. Here $m_u=313$
MeV, $g_{ch}=2.621$ are used. The three values for $E_0$ and $r_{\rm
rms}$ correspond to $m_b=4720$ MeV, 5100 MeV and 5259 MeV in order.
The first (second) $E_0$ and $r_{\rm rms}$ correspond to
$\Lambda$=1100 (1500) MeV. }\label{E0b-vec}
\end{table}

Up till now, we considered only neutral components of the system. In
Refs. \cite{liu-hqmole} and \cite{thomas}, the authors studied the
case with symmetric wave function case (i.e.
$a_0=a_1=\frac{1}{\sqrt2}$ in Eq. \ref{Xwave-ful}) and they found
the coupling to charged components is important. We also present the
numerical results for this case in Table \ref{E0c-sym} and
\ref{E0b-sym} which support the result that the channel coupling
should be considered in studying X(3872).

\begin{table}[htb]
\centering
\begin{tabular}{ccccc}\hline
    & $\chi$QM & \multicolumn{2}{c}{Ex. $\chi$QM } \\
    &                 & $f_{chv}/g_{chv}=0$ & $f_{chv}/g_{chv}=2/3$
\\\hline
$b_u$ (fm)& 0.5 & 0.45 & 0.45\\
$m_\sigma$ (MeV)& 595 & 535 & 547\\
$g_{chv}$ &     & 2.351 & 1.972 \\
$E_0$ (MeV)&$\times$& 12.1/14.2/19.3&4.5/6.0/9.7\\
$r_{\rm rms}$ (fm)&$\times$& 1.3/1.3/1.1&1.6/1.5/1.3 \\\hline
$E_0$ (MeV)&$\times$&16.3/18.6/24.5&6.9/8.6/13.0\\
$r_{\rm rms}$ (fm)&$\times$&1.2/1.2/1.0&1.5/1.4/1.2
\\\hline
\end{tabular}
\caption{Numerical results for the hidden charm case with the
symmetric wave function. Here $m_u=313$ MeV, $g_{ch}=2.621$ are
used. The three values for $E_0$ and $r_{\rm rms}$ correspond to
$m_c=1430$ MeV, 1550 MeV and 1870 MeV in order. The first (second)
set of $E_0$ and $r_{\rm rms}$ corresponds to $\Lambda$=1100 (1500)
MeV. $\times$ indicates the system is unbound.}\label{E0c-sym}
\end{table}

\begin{table}[htb]
\centering
\begin{tabular}{ccccc}\hline
    & $\chi$QM & \multicolumn{2}{c}{Ex. $\chi$QM } \\
    &                 & $f_{chv}/g_{chv}=0$ & $f_{chv}/g_{chv}=2/3$
\\\hline
$b_u$ (fm)& 0.5 & 0.45 & 0.45\\
$m_\sigma$ (MeV)& 595 & 535 & 547\\
$g_{chv}$ &     & 2.351 & 1.972 \\
$E_0$ (MeV)&12.4/13.8/14.4&47.6/50.0/50.9&32.0/34.0/34.8\\
$r_{\rm rms}$ (fm)&0.9/0.9/0.9&0.7/0.7/0.7&0.8/0.8/0.8 \\
\hline
$E_0$ (MeV)&14.6/16.2/16.8&56.4/59.0/60.0&38.3/40.5/41.4\\
$r_{\rm rms}$ (fm)&0.9/0.9/0.8&0.7/0.6/0.6&0.7/0.7/0.7
\\\hline
\end{tabular}
\caption{Numerical results for the hidden bottom case with the
symmetric wave function. Here $m_u=313$ MeV, $g_{ch}=2.621$ are
used. The three values for $E_0$ and $r_{\rm rms}$ correspond to
$m_b=4720$ MeV, 5100 MeV and 5259 MeV in order. The first (second)
set of $E_0$ and $r_{\rm rms}$ corresponds to $\Lambda$=1100 (1500)
MeV. }\label{E0b-sym}
\end{table}

\section{Summary and discussions}\label{summary}

In this work we have studied whether $D^0\bar{D}^{\ast0}$
($\bar{D}^0{D}^{\ast0}$) may form an S-wave molecule bound by the
$\pi$, $\sigma$, $\rho$ and $\omega$ exchange interactions in a
chiral quark model. These potentials are all attractive. By solving
the RGM equation, we failed to get a binding solution in this system
if we consider only $\pi$ and $\sigma$ contributions. When the
vector meson contributions are included, the existence of
$D^0\bar{D}^{\ast0}$ molecule seems to be possible. The coupling to
charged components is also important for a bound state.

When moving on to the heavier $B$ meson system, we obtain binding
state solutions. Our calculation favors the existence of an S-wave
$B\bar{B}^\ast$ ($\bar{B}{B}^\ast$) molecular state, which agrees
with the conclusion from Ref. \cite{liu-3872}. It will be very
interesting to search for such a bound state in the radiative decay
channel $X_B\to B^+B^- \gamma$ and the strong decay channel $X_B\to
\pi^+\pi^-\Upsilon$ in the future. Finding it may be possible at the
Tevatron or with the Large Hadron Collider beauty (LHCb) experiment
\cite{Hou}.

In the study of the deuteron, it was found that the tensor force
which mixes the S-wave and D-wave interactions is crucial in binding
the proton and the neutron. In an earlier calculation it was also
concluded that the tensor potential is very important in the mesonic
case \cite{Tornqvist}. In the present work, we did not consider
effects from the D-wave. Further study using the current approach
will be helpful to clarify whether this part can lead to a loosely
bound $D^0\bar{D}^{\ast0}$ ($\bar{D}^0{D}^{\ast0}$) state.

From the numerical values, we observe that vector meson
contributions are important in binding two color-singlet mesons.
However, the results rely on the vector coupling constants $g_{chv}$
and $f_{chv}$. Here we would like to mention that in our calculation
the parameters of light quark part are taken from Ref. \cite{ExCQM},
in which the calculated $NN$ scattering phase shifts and the binding
energy of deuteron are consistent with the experimental data. But
since the mechanism of the short range quark-quark interaction is
still an open problem, whether OGE or vector meson exchange is
dominate, or whether both of them are needed, one should be cautious
when making conclusions from these results.

In short summary, we have performed a dynamical calculation to
investigate whether the $D^0\bar{D}^{\ast0}$
($\bar{D}^0{D}^{\ast0}$) may form a molecule by considering the
$\pi$, $\sigma$, $\rho$ and $\omega$ exchange interactions. We could
not find an S-wave molecular state in this system in the chiral
quark model while its existence is not excluded in the extended
chiral quark model. More details of the dynamics should be
considered in further study of the X(3872). If it is really not a
molecule, the scheme of mixing a charmonium and a molecular state is
probably a way to solve the puzzles of the X(3872).

\section*{Acknowledgments}

YRL thanks Professor S.L. Zhu, Professor W.Z. Deng, Professor X.L.
Chen, Dr. F. Huang, Dr. X. Liu and C. Thomas for helpful
discussions. This project was supported by the National Natural
Science Foundation of China under Grants 10775146, 10805048, the
China Postdoctoral Science foundation (20070420526), and K.C. Wong
Education Foundation, Hong Kong.


\begin{thebibliography}{99}

\bibitem{Hmesons}E.S. Swanson, Phys. Rep. {\bf 429}, 243 (2006).
\bibitem{charmonium}M.B. Voloshin, Prog. Part. Nucl. Phys. {\bf 61}, 455 (2008); arXiv: 0711.4556 [hep-ph].
\bibitem{XYZ}S. Godfrey and S.L. Olsen, Ann. Rev. Nucl. Part. Sci. {\bf 58}, 51 (2008); arXiv: 0801.3867 [hep-ph].
\bibitem{newHadron}S.L. Zhu, Int. J. Mod. Phys. E {\bf 17}, 283 (2008); arXiv: 0707.2623 [hep-ph].


\bibitem{3872-first}Belle Collaboration, S.K. Choi et al., Phys. Rev. Lett. {\bf 91}, 262001 (2003).
\bibitem{3872-CDF}CDF Collaboration, D. Acosta et al., Phys. Rev. Lett. {\bf 93}, 072001 (2004).
\bibitem{3872-D0}D0 Collaboration, V.M. Abazov et al., Phys. Rev. Lett. {\bf 93}, 162002 (2003).
\bibitem{3872-BaBar}BaBar Collaboration, B. Aubert et al., Phys. Rev. D {\bf 71}, 071103
(2005).

\bibitem{PDG}W. M. Yao et al., Particle Data Group, J. Phys. G {\bf 33}, 1
(2006).

\bibitem{3872-angular-Belle}Belle Collaboration, K. Abe et al., arXiv: hep-ex/0505038.
\bibitem{3872-angular-CDF} CDF Collaboration, A. Abulencia et al., Phys. Rev. Lett {\bf 98}, 132002
(2007).

\bibitem{3872-charge}BaBar Collaboration, B. Aubert et al., Phys.
Rev. D {\bf 71}, 031501 (2005).

\bibitem{3872-rho-CDF}CDF Collaboration, A. Abulencia et al., Phys. Rev. Lett. {\bf 96}, 102002 (2006).

\bibitem{3872-gamma}Belle Collaboration, K. Abe et al., arXiv: hep-ex/0505037.

\bibitem{3872-gamma-Babar}Babar Collaboration, B. Aubert et al., Phys. Rev. {\bf D 74},
071101(R) (2006).

\bibitem{3872-Babar-ratio}BaBar Collaboration, B. Aubert et al.,
Phys. Rev. D {\bf 73}, 011101(R) (2006).

\bibitem{3875-Belle}Belle Collaboration, G. Gokhroo et al., Phys. Rev. Lett. {\bf 97}, 162002
(2006).
\bibitem{3875-BaBar}BaBar Collaboration, talk given by P.
Grenier in Moriond QCD 2007, 17-24 March, 2007,
http://moriond.in2p3.fr/QCD/2007/SundayAfternoon/\\Grenier.pdf; P.
Grenier, arXiv: 0705.2432 [hep-ex].


\bibitem{ccbar}T. Barnes, S. Godfrey, Phys. Rev. {\bf D 69}, 054008
(2004).
\bibitem{Mole-Close}F.E. Close, P.R. Page, Phys. Lett. {\bf B578}, 119
(2004).
\bibitem{Mole-Voloshin} M.B. Voloshin, Phys. Lett. {\bf B579}, 316 (2004); ibid {\bf B604}, 69 (2004).
\bibitem{Mole-Wong} C.Y.
Wong, Phys. Rev. C {\bf 69}, 055202 (2004).
\bibitem{Mole-Swanson} E.S. Swanson, Phys.
Lett. {\bf B588}, 189 (2004); ibid {\bf B598}, 197 (2004).
\bibitem{Mole-Torqvist} N.A. T\"ornqvist, Phys. Lett. {\bf B590}, 209 (2004).

\bibitem{3872-cusp}D.V. Bugg, Phys. Lett. {\bf B598}, 8 (2004).
\bibitem{3872-s-wave}J.L. Rosner, Phys. Rev. D {\bf 74}, 076006 (2006).
\bibitem{3872-hybrid}B.A. Li, Phys. Lett. {\bf B605}, 306 (2005).
\bibitem{3872-D}L. Maiani, F. Piccinini, A.D. Polosa, V. Riquer, Phys.
Rev. {\bf D 71}, 014028 (2005); L. Maiani, A.D. Polosa, and V.
Riquer, Phys. Rev. Lett. {\bf 99}, 182003 (2007)
\bibitem{3872-tetra}H. Hogaasen, J.M. Richard, P. Sorba, Phys. Rev. {\bf D 73}, 054013
(2006); D. Ebert, R.N. Faustov, V.O. Galkin, Phys. Lett. {\bf B
634}, 214 (2006); N. Barnea, J. Vijande, A. Valcarce, Phys. Rev.
{\bf D 73}, 054004 (2006); J. Vijande, E. Weissman, N. Barnea and A.
Valcarce, Phys. Rev. {\bf D 76}, 094022 (2007); Y. Cui, X.L. Chen,
W.Z. Deng, S.L. Zhu, High Energy Phys. Nucl. Phys. {\bf 31}, 7
(2007), arXiv: hep-ph/0607226; R.D. Matheus, S. Narison, M. Nielsen,
J.M. Richard, Phys. Rev. {\bf D 75}, 014005 (2007); T.W. Chiu, T.H.
Hsieh, Phys. Lett. {\bf B 646}, 95 (2007); K. Terasaki,
arXiv:0706.3944 [hep-ph], Prog. Theor. Phys. {\bf 118}, 821 (2007).

\bibitem{3872-glueball}K.K. Seth, Phys. Lett. {\bf B 612}, 1 (2005).

\bibitem{3872-res}D. Gamermann and E. Oset, Eur. Phys. J. A {\bf 33}, 119
(2007); arXiv: 0712.1758 [hep-ph].

\bibitem{EichtenLQ}E.J. Eichten, K. Lane, C. Quigg, Phys. Rev. {\bf D 69}, 094019
(2004).
\bibitem{suzuki}M. Suzuki, Phys. Rev. {\bf D 72}, 114013 (2005).
\bibitem{chao-3872}C. Meng, Y.J. Gao and K.T. Chao, arXiv:
hep-ph/0506222; C. Meng and K.T.Chao, Phys. Rev. {\bf D 75}, 114002
(2007).
\bibitem{Kalashinikova}Y.S. Kalashnikova, Phys. Rev. {\bf D 72}, 034010 (2005).
\bibitem{Pennington}M.R. Pennington and D.J. Wilson, Phys. Rev. {\bf
D 76}, 077502 (2007).

\bibitem{Okun}M.B. Voloshin and L.B. Okun, JETP Lett. {\bf 23}, 333 (1976).
\bibitem{RGG}A.De Rujula, H. Georgi and S.L. Glashow, Phys. Rev.
Lett. {\bf 38}, 317 (1977).
\bibitem{Tornqvist}N.A. T\"{o}rnqvist, Phys. Rev. Lett. {\bf 67}, 556
(1991); Z. Phys. C {\bf 61}, 525 (1994).

\bibitem{Petrov}M.T. AlFiky, F. Gabbiani, A.A. Petrov, Phys. Lett.
{\bf B 640}, 238 (2006).
\bibitem{Colangelo}P. Colangelo, F.De Fazio, S. Nicotri, Phys. Lett.
{\bf B 650}, 166 (2007).
\bibitem{Braaten-ratio}E. Braaten, M. Kusunoki, Phys. Rev. D {\bf 71}, 074005 (2005).
\bibitem{Braaten-3872}E. Braaten, M. Kusunoki, S. Nussinov, Phys.
Rev. Lett. {\bf 93}, 162001 (2004); E. Braaten, M. Kusunoki, Phys.
Rev. D {\bf 69}, 074005 (2004); Phys. Rev. D {\bf 69}, 114012
(2004); Phys. Rev. D {\bf 72}, 014012 (2005); Phys. Rev. D {\bf 72},
054022(2005); E. Braaten, Phys, Rev. D {\bf 73}, 011501(R) (2006);
Phys. Rev. D {\bf 77}, 034019 (2008) E. Braaten, M. Lu, Phys. Rev. D
{\bf 74}, 054020 (2006); Phys. Rev. D {\bf 76}, 094028 (2007); E.
Braaten, M. Lu, J. Lee, Phys. Rev. D {\bf 76}, 054010 (2007).
\bibitem{Voloshin}S. Dubynskiy, M.B. Voloshin, Phys. Rev. D {\bf 74},
094017 (2006); M.B. Voloshin, Phys. Rev. D {\bf 76}, 014007 (2007).
\bibitem{Kolck}S. Fleming, M. Kusunoki, T. Mehen, and U. van Kolck, Phys. Rev.
D {\bf 76}, 034006 (2007).
\bibitem{Hanhart}C. Hanhart, Y.S. Kalashnikova, A.E. Kudryavtsev, and A.V.
Nefediev, Phys. Rev. {\bf D 76}, 034007 (2007).
\bibitem{Ma}G.Y. Chen, J.P. Ma, Phys. Rev. D {\bf 77}, 097501, (2008).
\bibitem{Dong}Y. Dong, A. Faessler, T. Gutsche, V.E. Lyubovitskij, Phys. Rev. D {\bf 77}, 094013 (2008).



\bibitem{liu-3872}Y.R. Liu, X. Liu, W.Z. Deng and S.L. Zhu, Eur. Phys. J. C {\bf 56}, 63 (2008), arXiv:0801.3540
[hep-ph].
\bibitem{liu-4430}X. Liu, Y.R. Liu, W.Z. Deng and S.L. Zhu,
Phys. Rev. D {\bf 77}, 034003 (2008); arXiv:0803.1295 [hep-ph]; X.
Liu, Y.R. Liu, W.Z. Deng, arXiv:0802.3157 [hep-ph].



\bibitem{Kamimura}M. Kamimura, Suppl. Prog. Theor. Phys. {\bf 62}, 236
(1977).
\bibitem{Oka}M. Oka and K. Yazaki, Prog. Theor. Phys. {\bf 66}, 556 (1981).
\bibitem{SU3CQM}Z.Y. Zhang. Y.W. Yu, P.N. Shen, L.R. Dai, A. Faessler, and
U. Straub, Nucl. Phys. {\bf A625}, 59 (1997).
\bibitem{ExCQM}L.R. Dai, Z.Y. Zhang, Y.W. Yu and P. Wang, Nucl. Phys. {\bf A727}, 321 (2003).
\bibitem{BMsystem}F. Huang, Z.Y. Zhang and Y.W. Yu, Phys. Rev. C {\bf 70},
044004 (2004); ibid. {\bf 72}, 065208 (2005); ibid. {\bf 73}, 025207
(2006); High Energy Phys. Nucl. Phys. {\bf 29}, 948 (2005); Commun.
Theor. Phys. {\bf 44}, 665 (2005); F. Huang and Z.Y. Zhang, Phys.
Rev. C {\bf 70}, 064004 (2004); ibid. {\bf 72}, 024003 (2005); ibid.
{\bf 72}, 068201 (2005); F. Huang, D. Zhang, Z.Y. Zhang and Y.W. Yu,
ibid. {\bf 71}, 064001 (2005); F. Huang, W.L. Wang, Z.Y. Zhang, Y.W.
Yu, ibid. {\bf 76}, 018201 (2007); W.L. Wang, F. Huang, Z.Y. Zhang,
Y.W. Yu, and F. Liu, Eur. Phys. J. A {\bf 32}, 293 (2007).
\bibitem{BBbarsys}D. Zhang, F. Huang, L.R. Dai, Y.W. Yu and Z.Y.
Zhang, Phys. Rev. C {\bf 75}, 024001 (2007); H.R. Pang, J.L. Ping,
F. Wang, Chin. Phys. Lett. {\bf 25}, 3192 (2008).

\bibitem{zzz}H.X. Zhang, W.L. Wang, Y.B. Dai, Z.Y. Zhang,
hep-ph/0607207.

\bibitem{Vijande}J. Vijande, H. Garcilazo, A. Valcarce, and F. Fernandez, Phys. Rev. D {\bf 70}, 054022 (2004)
\bibitem{Semay}B. Silvestre-Brac and C. Semay, Z. Phys. C {\bf 57}, 273 (1993).

\bibitem{zhangd}D. Zhang, F. Huang, Z.Y. Zhang, Y.W. Yu, Nucl. Phys.
{\bf A756}, 215 (2005).

\bibitem{Riska}D.O. Riska, G.E. Brown, Nucl. Phys. {\bf A679}, 577
(2001).

\bibitem{zzz2}H.X. Zhang, M. Zhang, Z.Y. Zhang,
Chin. Phys. Lett. {\bf 24}, 2533 (2007); M. Zhang, H.X. Zhang, Z.Y.
Zhang, Commun. Theor. Phys. {\bf 50}, 437 (2008), arXiv:0711.1029
[nucl-th].

\bibitem{Vijande2}J. Vijande, F. Fernandez and A. Valcarce, J. Phys. G {\bf 31},
481 (2005).

\bibitem{myerr}The qualitative properties for meson exchange potentials in
different frameworks should be consistent. When comparing the
present result with the potentials at hadron level, we found
inconsistencies. In fact, we missed a minus sign when deriving the
sigma exchange potential in equation (13) of reference
\cite{liu-3872}. Therefore that potential should also be attractive
(see also reference \cite{liu-hqmole}). The sigma exchange terms in
equations. (15), (18), and (22) also change sign. Fortunately, the
total potential and the numerical results change little and the
final conclusion is not affected. As for the vector meson exchange
potentials for $D\bar{D}^*$ system, we obtained attractive forces at
hadron level in reference \cite{liu-hqmole}. After carefully
checking the calculation in the present framework, we confirmed
those signs for $\rho$ and $\omega$ exchange potentials.

\bibitem{liu-hqmole}Xiang Liu, Zhi-Gang Luo, Yan-Rui Liu, Shi-Lin
Zhu, arXiv: 0808.0073 [hep-ph].

\bibitem{thomas}C.E. Thomas, F.E. Close, Phys. Rev. D {\bf 78}, 034007, (2008), arXiv: 0805.3653 [hep-ph].

\bibitem{Hou}W.S. Hou, Phys. Rev. D {\bf 74}, 017504 (2006).


\end{thebibliography}
\end{document}